\documentclass{midl} 

\usepackage{mwe} 
\usepackage{setspace}
\usepackage{bibspacing}
\newcommand{\squeezeup}{\vspace{-5mm}}
\usepackage[font=small,labelfont=bf]{caption}

\usepackage{graphicx}
\usepackage{float}
\jmlrproceedings{MIDL}{Medical Imaging with Deep Learning}
\jmlrpages{}
\jmlryear{2022}

\jmlrworkshop{Short Paper -- MIDL 2022}
\editors{Accepted for publication at MIDL 2022}

\title[Predicting Thrombectomy Recanalization from Admission CT Imaging Using Deep Learning Models]{Predicting Thrombectomy Recanalization from CT Imaging Using Deep Learning Models}

\midlauthor{\Name{Haoyue Zhang\midljointauthortext{Haoyue Zhang and Jennifer S Polson contributed equally.}\nametag{$^{1,2}$}} \Email{HaoyueZhang@mednet.ucla.edu}\\
\Name{Jennifer S Polson\midlotherjointauthor\nametag{$^{1,2}$}} \Email{jpolson@g.ucla.edu}\\
\Name{Eric J Yang\nametag{$^{1}$}}\Email{EricJYang@ucla.edu}\\
\Name{Kambiz Nael\nametag{$^{3}$}} \Email{kanael@mednet.ucla.edu}\\
\Name{William Speier\nametag{$^{1,3}$}} \Email{Speier@ucla.edu}\\
\Name{Corey W. Arnold\nametag{$^{1,2,3,4}$}} \Email{cwarnold@ucla.edu}\\
\addr $^{1}$ Computational Diagnostics Lab, University of California, Los Angeles, CA 90024 USA\\\
\addr $^{2}$ Department of Bioengineering, University of California, Los Angeles, CA 90024 USA\\\
\addr $^{3}$ Department of Radiology, University of California, Los Angeles, CA 90024, USA\\\
\addr $^{4}$ Department of Pathology, University of California, Los Angeles, CA 90024, USA
}
\begin{document}
\maketitle
\squeezeup
\begin{abstract}
 For acute ischemic stroke (AIS) patients with large vessel occlusions, clinicians must decide if the benefit of mechanical thrombectomy (MTB) outweighs the risks and potential complications following an invasive procedure. Pre-treatment computed tomography (CT) and angiography (CTA) are widely used to characterize occlusions in the brain vasculature. If a patient is deemed eligible, a modified treatment in cerebral ischemia (mTICI) score will be used to grade how well blood flow is reestablished throughout and following the MTB procedure. An estimation of the likelihood of successful recanalization can support treatment decision-making. In this study, we proposed a fully automated prediction of a patient's recanalization score using pre-treatment CT and CTA imaging. We designed a spatial cross attention network (SCANet) that utilizes vision transformers to localize to pertinent slices and brain regions. Our top model achieved an average cross-validated ROC-AUC of 77.33 $\pm$ 3.9\%. This is a promising result that supports future applications of deep learning on CT and CTA for the identification of eligible AIS patients for MTB.
\end{abstract}

\begin{keywords}
Acute Ischemic Stroke, Computed Tomography, Deep Learning, Vision Transformers
\end{keywords}

\section{Introduction}
Stroke is the fifth leading cause of death and the leading cause of long-term disability; of the 795,000 new and recurrent strokes each year, acute ischemic stroke (AIS) accounts for 87\% of cases \cite{Tsao:Stroke}. Mechanical thrombectomy (MTB) is the leading treatment for patients with clots in large blood vessels. In this procedure, a blood clot is surgically removed from an artery to achieve recanalization, i.e., restored blood flow. As a standard measurement for recanalization achieved, a modified treatment in cerebral ischemia (mTICI) score \cite{Tomsick:mTICI} is assigned to patients post-treatment. This post-treatment score is clinically significant, as it has been shown that favorable scores, i.e., mTICI 2c or greater, are associated with better clinical outcomes in the long term \cite{Chamorro:Reperfusion}. Unfavorable scores (mTICI less than 2c) indicate that the treatment did not effectively clear the blood vessel. Imaging has been identified as one modality to illustrate patient physiology that could influence the likelihood of a successful MTB procedure. Predicting final mTICI score prior to a procedure can provide doctors and with more more information when considering treatment options. Deep learning has been shown to leverage the amount of detail in images to improve prediction accuracy \cite{LeCun:Deep_Learning}. Current literature presents models that perform semi-automated prediction of mTICI score based on pre-treatment CT imaging, with inconsistent performance \cite{Hilbert:Semi-automated,siddiqui2021quantitative}. We propose a fully automated model that uses both CT and CTA images to predict mTICI score post-treatment, incorporating attention modules into a deep learning network to effectively localize to informative stroke regions without requiring manual segmentation. 

\section{Data and Methods}
The cohort used for this study comprises patients treated from 2012-2019. A patient was included in the cohort if they had CT and CTA imaging, underwent thrombectomy for stroke, and were assigned an mTICI score post-MTB. Of the 254 eligible patients, 69 patients were excluded due to missing either CT or CTA series, and 8 were excluded to due unclear stroke location, leaving 177 patients total. The dataset matched demographic distributions seen in other stroke studies, and the target labels were approximately balanced. Patient images were processed using a previously published pipeline adapted for CT, which included brain extraction and registration to a CT template in MNI space \cite{zhang2021intra}.
\begin{figure}[t]
    \centering
    \includegraphics[width=\textwidth]{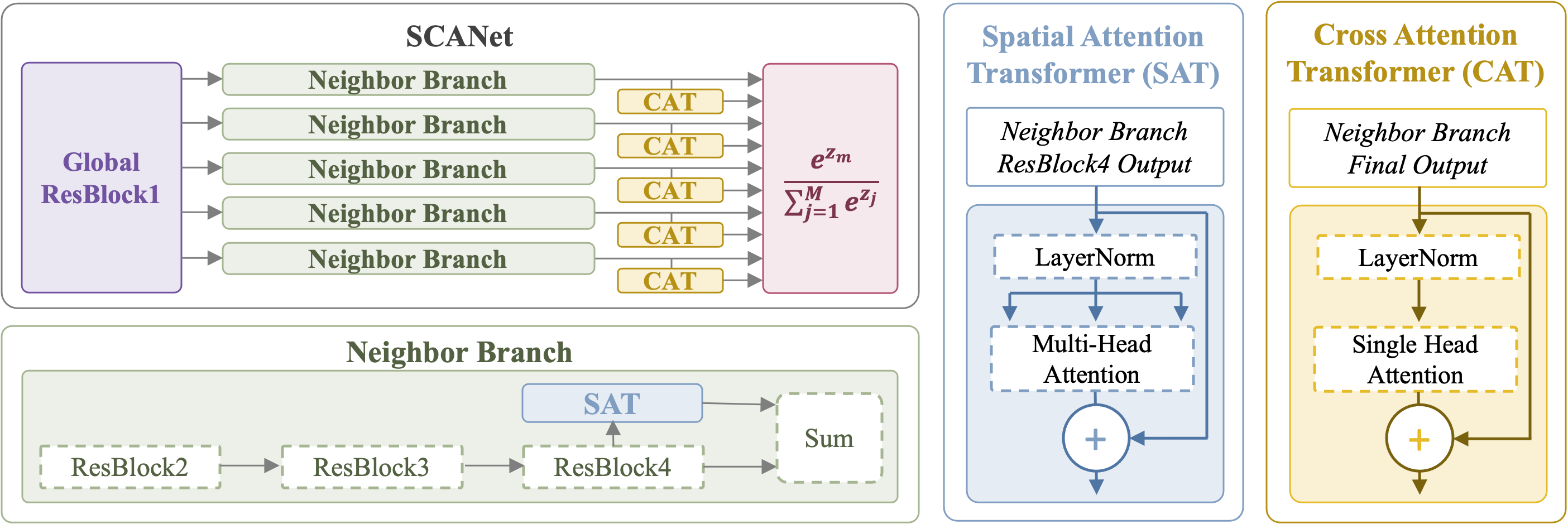}
    \squeezeup
    \caption{Overview of the proposed deep learning architecture. The figure details the entire architecture, neighborhood branch modules, spatial attention transformer module (SAT), and cross attention transformer module (CAT).}
    \label{fig:arch}
    \squeezeup
\end{figure}

Utilizing a ResNet backbone, CT and CTA images were used as a slice-wise input (input size 26x224x224) for a global 2D convolutional block, chosen because of the large slice thickness found in stroke protocols \cite{he2016deep}. The outputs from neighboring slices were then fed into  ResNet34-based branches that shared weights across slice neighborhoods. The model leveraged two versions of transformer attention modules. A spatial attention transformer (SAT) utilized multi-head attention on each slice to focus on salient regions \cite{dosovitskiy2020vit}. Within each neighborhood branch, a cross attention transformer (CAT) identified important slices. Finally, the branch outputs were subjected to a weighted softmax layer to ultimately generate binary predictions. The architecture and modules are summarized in Figure \ref{fig:arch}. The model was trained for 200 epochs with early stopping, using the Adam optimizer with weight decay, a batch size of 12, and a learning rate of 0.0001.


\begin{table}[htbp]
\footnotesize
\begin{tabular}{lccccc}
\hline
 \textbf{Model} & \textbf{ROC-AUC} & \textbf{Accuracy} & \textbf{Precision} & \textbf{Sensitivity} & \textbf{Specificity} \\ \hline
\textbf{Siddiqui et al.} & 0.717 & -- & -- & -- & -- \\
\textbf{Hilber et al.} & 0.65  $\pm$  0.10 & -- & -- & -- & -- \\
\textbf{Radiomics} & 0.6859 $\pm$ 0.043 & 0.6877 $\pm$ 0.039 & 0.6417 $\pm$ 0.068  & 0.7425 $\pm$ 0.123 & 0.6421 $\pm$ 0.126 \\
\textbf{ResNet34} & 0.5840 $\pm$ 0.036 & 0.5656 $\pm$ 0.046 & 0.5410 $\pm$ 0.067 & 0.8500 $\pm$ 0.300 & 0.3253 $\pm$ 0.296\\
\textbf{SCANet} & 0.7732 $\pm$ 0.039 & 0.7523 $\pm$ 0.042 & 0.7424 $\pm$ 0.145 & 0.8250 $\pm$ 0.174 & 0.6905 $\pm$ 0.215 \\ \hline

\end{tabular}
\caption{Performance of our current model benchmarked against results from literature as well as previously published models applied to this cohort \cite{Hilbert:Semi-automated,siddiqui2021quantitative}.}
\label{tab:results}
\squeezeup
\end{table}

\section{Results and Discussion}
The results of our experiments are summarized in Table \ref{tab:results}. The average ROC-AUC achieved by SCANet was 0.7732 $\pm$ 0.039. This is a significant improvement over the previously published fully automatic deep learning model\cite{siddiqui2021quantitative}. Our method also demonstrates higher and more robust performance metrics than the state-of-the-art model requiring manual clot segmentation  \cite{Hilbert:Semi-automated}. In addition to the literature benchmarks, SCANet performs better than a radiomics-based model and standard deep learning architecture when trained on the same cohort \cite{radiomics2021}.\

Clinicians decide to perform MTB based on likelihood of successful recanalization, but it is unknown what factors underlie MTB responses. Clinical images such as CT and CTA contain valuable information to predict procedure outcome, and deep learning models have the capability to learn representations from highly dimensional imaging data. This study sought to predict final MTB recanalization in a fully automatic manner, leveraging recent advances in vision transformers to localize to the stroke region. We showed that our proposed model outperforms prior fully- and semi-automated machine and deep learning models. The primary limitation of our study is the small sample size, which precludes more robust validation. A few future directions include experimenting on a larger dataset across several institutions, optimizing the preprocessing pipeline to more effectively preserve high resolution CTA, and correlation of the immediate treatment response with long-term outcomes. These steps can produce a model that more accurately predicts MTB recanalization, in turn helping doctors and patients in the treatment decision process.

\squeezeup
{\tiny
\let\oldbibliography\thebibliography
\bibliography{midl-bib}}
\squeezeup
\end{document}